\documentclass[preprint,showpacs,superscriptaddress,floatfix,prl]{revtex4}

\usepackage[dvips]{graphicx}
\usepackage{epsfig}
\usepackage{tabularx}

\newcommand{\celsius}{$^\circ$C}

\begin{document}

\title{Experimental measurement of photothermal effect in Fabry--Perot cavities}

\author{M.~De~Rosa}
\email[Corresponding author: ]{derosa@lens.unifi.it}

\affiliation{Dipartimento di Fisica, Universit\`a di Firenze, INFN, Sezione di Firenze, and LENS \\ Via Sansone 1, I-50019 Sesto Fiorentino (FI), Italy}

\author{L.~Conti}
\affiliation{Dipartimento di Fisica, Universit\`a di Padova, and INFN, Sezione di Padova\\ Via Marzolo 8, I-35131 Padova, Italy}

\author{M.~Cerdonio}
\affiliation{Dipartimento di Fisica, Universit\`a di Padova, and INFN, Sezione di Padova\\ Via Marzolo 8, I-35131 Padova, Italy}

\author{M.~Pinard}
\affiliation{Laboratoire Kastler Brossel, 4 place Jussieu, F75252 Paris, France}

\author{F.~Marin}
\affiliation{Dipartimento di Fisica, Universit\`a di Firenze, INFN, Sezione di Firenze, and LENS \\ Via Sansone 1, I-50019 Sesto Fiorentino (FI), Italy}
\date{\today}
\begin{abstract}
We report the experimental observation of the photothermal effect. The measurements are performed by modulating  the laser power absorbed by the mirrors of two high-finesse Fabry-Perot cavities. The results are very well described by a recently proposed theoretical model [M.~Cerdonio, L.~Conti, A.~Heidmann and M.~Pinard, Phys. Rev. D {\bf 63} (2001) 082003], confirming the correctness of such calculations. Our observations and quantitative characterization of the photothermal effect demonstrate its critical importance for interferometric displacement measurements towards the quantum limit, as those necessary for gravitational wave detection.
\end{abstract}


\maketitle

The study of thermal noise in optical components has a great interest both for the basic physics of the involved phenomena and for its implications in determining the fundamental limits of displacement measurements. In particular, the surface of the mirrors undergoes fluctuations which can depend on several dissipation mechanisms. Brownian noise, or volume fluctuations, is generally related to homogeneous dissipation inside the mass of the mirrors, and is studied long since~\cite{Saulson,Levin}. However, other dissipative mechanisms can induce surface fluctuations which can exceed the Brownian ones. Recently several authors examined the effects of temperature fluctuations which translate into surface fluctuations via the thermal expansion coefficient~\cite{Braginsky1,Liu,Cerdonio}; in particular they considered thermodynamic fluctuations of temperature of the  mirrors and fluctuations caused by randomly absorbed photons ({\em photothermal effect}). 
Thermodynamic fluctuations of temperature are an intrinsic effect of the mirror material (substrate and reflecting coating) and are independent of the radiation power on the mirror.
On the contrary, photothermal effect, beside the dependence on the thermal property of the mirror materials, depends on the power impinging on the mirror surface as well.
Indeed, when light reflects off the surface of a mirror, a fraction of it is absorbed at the surface causing a local temperature change and a local deformation of the mirror surface. 
A fundamental limit to the position fluctuations due to photothermal effect is given by the standard quantum noise of the laser.

Accurate studies of thermal effects are of crucial importance for the applications to gravitational wave detectors. The sensitivity of long arm Michelson interferometers~\cite{interferometri} depends critically from such effects, and also for massive detectors the most recently studied readout schemes rely on opto-mechanical techniques~\cite{Derosa}.
In particular, Braginsky {\em et al.}~\cite{Braginsky1} pointed out that the performance of advanced interferometric GW detectors, such as LIGOII~\cite{ligo2}, might be limited by thermodynamic fluctuations of temperature and by photothermal effect.
They derived an approximated expression for the spectral power density of photothermal noise which is valid for time scales smaller than the thermal relaxation time $\tau_c$ (adiabatic limit), i.e., for angular frequencies $\omega$ larger than 
\begin{equation}
\omega_c = \frac{1}{\tau_c} = \frac{\kappa}{\rho \, C \, r_0^2} \, ,
\label{omega}
\end{equation}
where $\kappa$ is the thermal conductivity of the mirror, $\rho$ is the density, $C$ is the specific heat capacity and $r_0$ is the beam radius (assuming a Gaussian transverse profile), defined as the radius at which the power drops to $1/e$ of its central value.
More recently the calculation has been extended  by Cerdonio {\em et al.}~\cite{Cerdonio} to angular frequencies $\omega$ smaller than the adiabatic limit, defined by Eq.~(\ref{omega}).
They predicted an expression for the power spectral density of the mirror surface position $S_{\rm PT}$ due to the photothermal noise, as
\begin{equation} 
S_{\rm PT}(\omega) = \frac{2}{\pi^2} \, (1+\sigma)^2 \, \frac{\alpha^2}{\kappa^2} \, S_{\rm abs}(\omega) \, K(\Omega) \, ,
\label{eq:Sxx}
\end{equation} 
where $\sigma$ is the Poisson ratio, $\alpha$ is the thermal expansion coefficient, $S_{\rm abs}(\omega)$ is the spectral power noise of the absorbed light\footnote{The factor 2 on the right-hand side of Eq.~\ref{eq:Sxx} is due to a different definition adopted in Ref.~\cite{Cerdonio} for the two power spectral density: $S_{\rm PT}(\omega)$ is a two-sided spectral density, while $S_{\rm abs}(\omega)$ is a one-sided spectral density.}, and 
\begin{equation}
K(\Omega) = \left| \frac{1}{\pi} \int_0^\infty \!\!\!\! du \int_{-\infty}^\infty \!\!\! dv \;\frac{u^2 e^{-u^2/2}}{(u^2+v^2)(u^2+v^2+i \Omega)}\right|^2\label{eq:integral}
\end{equation}
is a shape function of the adimensional variable $\Omega=\omega/\omega_c$. Assuming a flat $S_{\rm abs}$, as in the case of shot noise, this model predicts that for $\Omega \gg 1$ $S_{\rm PT}$ is proportional to $\Omega^{-2}$, whereas for $\Omega \ll 1$ it is almost constant.

There exist some experimental evidence that the mirror heating changes the lenght of optical cavities. In Ref.~\cite{FEL} the authors describe the measurement of the output power of a CO$_2$ laser at the turn-on and find a good agreement with simulations of the laser optical cavity where the mirror surface is deformed by heating. In a different experiment, An et al.~\cite{MIT} measure the transmission through a high-finesse Fabry--Perot when the laser is rapidly swept. They find an hysteretic behavior well explained by considering the mirror heating. Though their observations represent an indication of photothermal effect, their measurements are limited to the adiabatic limit defined by Eq.~(\ref{omega}) and can hardly be compared with accurate models of the frequency response. 
So far the mentioned theoretical predictions have not yet been verified by experimental investigations, which are indeed the subject of the present work.

In this paper we report for the first time on the experimental measurement of the frequency behavior of the photothermal effect: the measurements have been performed over a frequency range not limited to the adiabatic regime, 
so they can be compared with the prediction of the extended model proposed in Ref.~\cite{Cerdonio}. 

\begin{figure}[t]
\begin{center}
\includegraphics[width=\columnwidth]{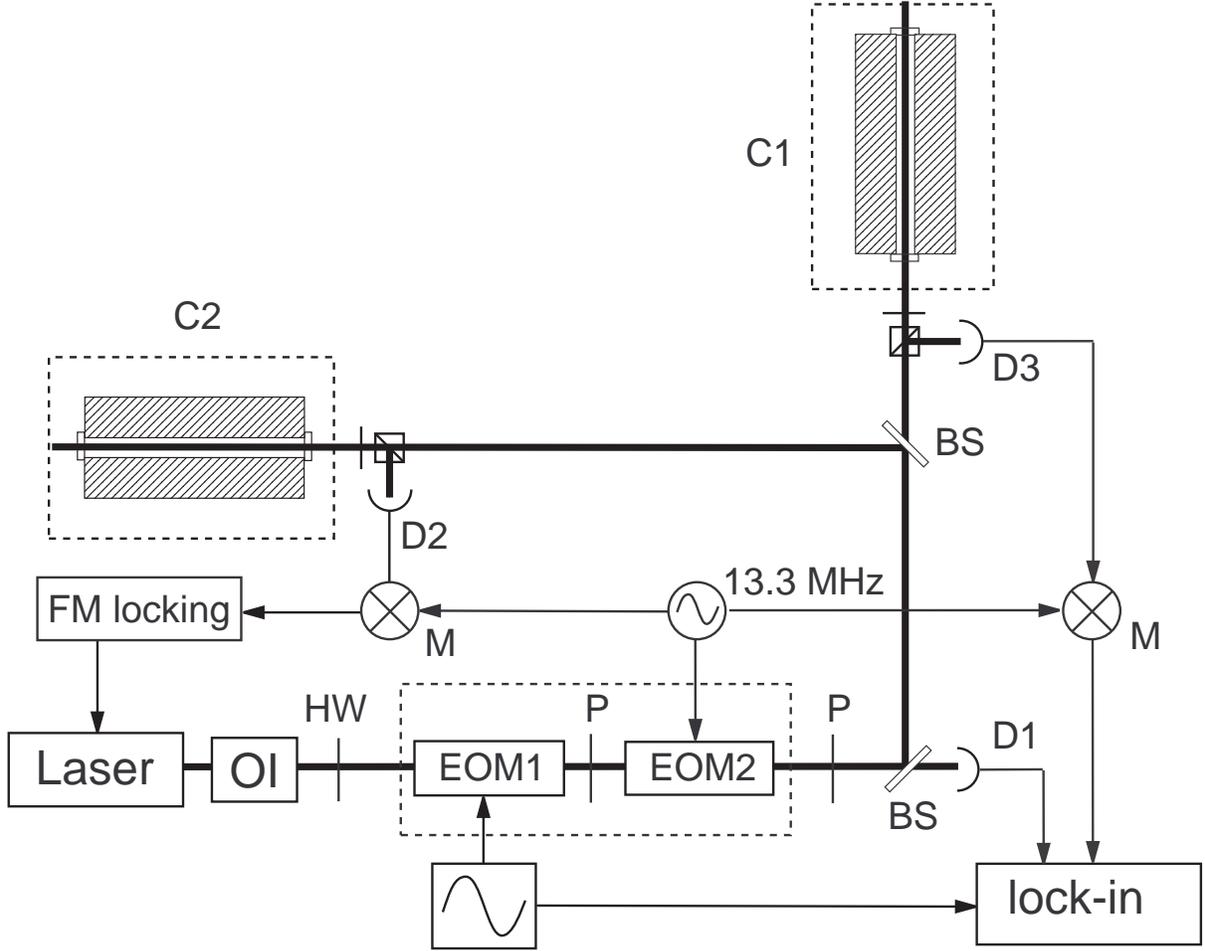} 
\end{center}
\caption{Scheme of the experimental setup: C\#, Fabry--Perot cavity; OI, optical isolator; HW, half-wave plate; EOM\#, electro-optic modulator; P, polarizer; BS, beam-splitter; D\#, photo-detector; M, mixer. Dashed lines mark out active thermal stabilization.}
\label{setup}
\end{figure}
The mirrors are part of two high-finesse Fabry-Perot interferometers whose length variations are measured through phase-sensitive detection of resonant laser radiation. 
Our experimental apparatus was originally conceived to realize and test an optical readout for the gravitational wave bar detector AURIGA~\cite{Conti-RSI}; however we were able to observe the effect by actively modulating the intracavity light power, with minimal modifications of the setup.

The experimental setup is sketched in Fig.~\ref{setup}. The laser system and the detection are described in details in Ref.~\cite{Conti-AO}. The light source is a cw Nd:YAG laser based on a non-planar ring oscillator geometry (Lightwave, mod. 126-1064-100), emitting 100~mW at 1064~nm.
A non-resonant electro-optic modulator (EOM1) followed by a polarizer acts as an amplitude modulator for the transmitted laser radiation: the EOM1 has one crystal axis rotated by an angle of about 0.12~rad with respect to the input (vertical) polarization while the polarizer restores the vertical polarization. Applying a voltage to the EOM1 changes the ellipticity of the beam exiting EOM1 and thus the amplitude of light transmitted by the polarizer. 
By properly setting the temperature of the EOM1 we can reach and maintain a working point for where the variation of the transmitted light is proportional to the applied voltage.
With no signal fed to the EOM1 the transmission after the polarizer is about 92$\%$. This setup is part of a power stabilization loop described elsewhere~\cite{Conti-AO} but in this work it is used merely for power modulation. A second, resonant electro-optic modulator (EOM2) phase modulates the laser output at 13.3~MHz. The two EOMs are enclosed in an aluminum box whose temperature is actively stabilized at $\sim 35$\celsius\ within 0.04\celsius. A polarizer after EOM2, parallel to the previous one, removes any ellipticity that might originate in the phase modulation. The beam then crosses a first 50$\%$ beam splitter: the transmitted beam is detected by a photodetector (D1) for power monitoring; the reflected beam is successively divided in two by a second 50$\%$ beam splitter. Each of the outgoing beams, of about 10~mW, passes an optical circulator formed by a polarizing beam-splitter and a quarter-of-wave plate and is then directed towards a Fabry-Perot cavity (C1 and C2 in Fig.~\ref{setup}). The reflected beams are detected by photodiodes and the signals are demodulated and filtered, according to the FM sideband technique~\cite{Drever}.

The two Fabry-Perot cavities are made by mirrors coated on fused silica substrates, which are optically contacted to a 20~cm long Zerodur spacer. Each cavity is placed on a double stage mechanical suspension which isolates from mechanical noise and it is enclosed in a vacuum chamber evacuated by an ion pump. The cavity temperature is actively stabilized within 1~mK. By varying the cavity temperature between 20\celsius\ and 100\celsius, it is possible to change the length of the cavity by roughly 0.5~$\mu$m, corresponding to one Free Spectral Range (FSR), and consequently to tune coarsely the optical resonances of the cavities. Fine adjustments of the frequency of C1 is accomplished by a system of piezoelectric actuators (PZT) and invar rods which compresses the Zerodur spacer. Main figures of the two cavities are summarized in Table~\ref{tab1}. These cavities are designed as length reference for the AURIGA optical readout which is optimized for a frequency range around 1~kHz where it is characterized by a relative displacement noise of few $10^{-18}$~m/$\sqrt{\rm Hz}$. This noise increases as $1/f$ for lower frequencies

The FM sideband signal from C2 is used as discriminator for frequency locking the laser. Once two resonance peaks of the two cavities have been coarsely superimposed, the laser frequency is locked to a resonance peak of C2 with a servo loop which has a unit gain frequency of 30~kHz and a gain of 120~dB around 1~kHz. C1 is then finely tuned by acting on the PZT. The FM sideband signal from this cavity is used to monitor the relative frequency change and thus relative variations of the cavity lengths. 
\begin{table}[t]
\caption{Parameters of the two cavities. We report the beam waist for each mirror, as usually defined in Gaussian optics, which differ from the beam radius $r_0$, as defined in the text, by a factor $\sqrt{2}$.}
\label{tab1}
\begin{center}
\begin{tabular}{lcccc}
\hline \hline 
Cavity & \multicolumn{2}{c}{C1}  & \multicolumn{2}{c}{C2} \\
\hline
Length (m) & \multicolumn{2}{c}{0.2} & \multicolumn{2}{c}{0.2} \\
FSR (MHz) & \multicolumn{2}{c}{750} & \multicolumn{2}{c}{750} \\
Finesse   & \multicolumn{2}{c}{16000} & \multicolumn{2}{c}{36000} \\
Intracavity power (W)& \multicolumn{2}{c}{20} & \multicolumn{2}{c}{240}\\
\hline
Mirror & in & out & in & out \\
\hline
Radius of curvature (m)   & 1 & 1 & $\infty$ & 1 \\
Waist (mm) & 0.336 & 0.336 & 0.368 & 0.412 \\
\hline \hline 
\end{tabular}
\end{center}
\end{table}

To observe the photothermal effect we apply a sinusoidal voltage at frequency $\nu_{\rm m}=\omega_m/2\pi$ to EOM1. The power incident on both cavities is thus modulated around an average value $P_0$, according to 
\[P=P_0 \, + \, P_m \, \sin \omega_m t ,
\]
where $P_m$ is the modulation depth. Consequently the light power absorbed by the mirrors is modulated and a cavity length change is originated due to the non-zero thermal expansion coefficient of the fused silica substrates.
We simultaneously acquire the signal from D1, to monitor the power modulation, and the FM sideband signal from C1; these two signals are successively phase-detected at the frequency $\nu_{\rm m}$ with a digital, numerical lock-in. The modulation frequency $\nu_{\rm m}$ ranges between $5 \times 10^{-3}$~Hz and $5 \times 10^2$~Hz, with constant modulation depth ($P_m/P_0 \sim 0.01)$.
Measurements have also been made by fixing the modulation frequency at $\nu_{\rm m}=0.05$~Hz and varying the modulation depth.

If the two cavities were exactly identical we would observe no effects, since each cavity would experience the same length variation. However, as shown in Table~\ref{tab1}, due to different mirror losses the two cavities have different finesse and intracavity power, whereas the impinging laser power is the same.
In particular, the power circulating into C2 is one order of magnitude larger than that into C1.  Hence, we expect a differential length variation of amplitude $\delta l$ between the two cavities and, as will be explained later, we can reasonably assume the observed effect is mainly due to C2.

\begin{figure}[t]
\begin{center}
\includegraphics*[bbllx=80bp,bblly=45bp,bburx=710bp,bbury=520bp,width=\columnwidth]{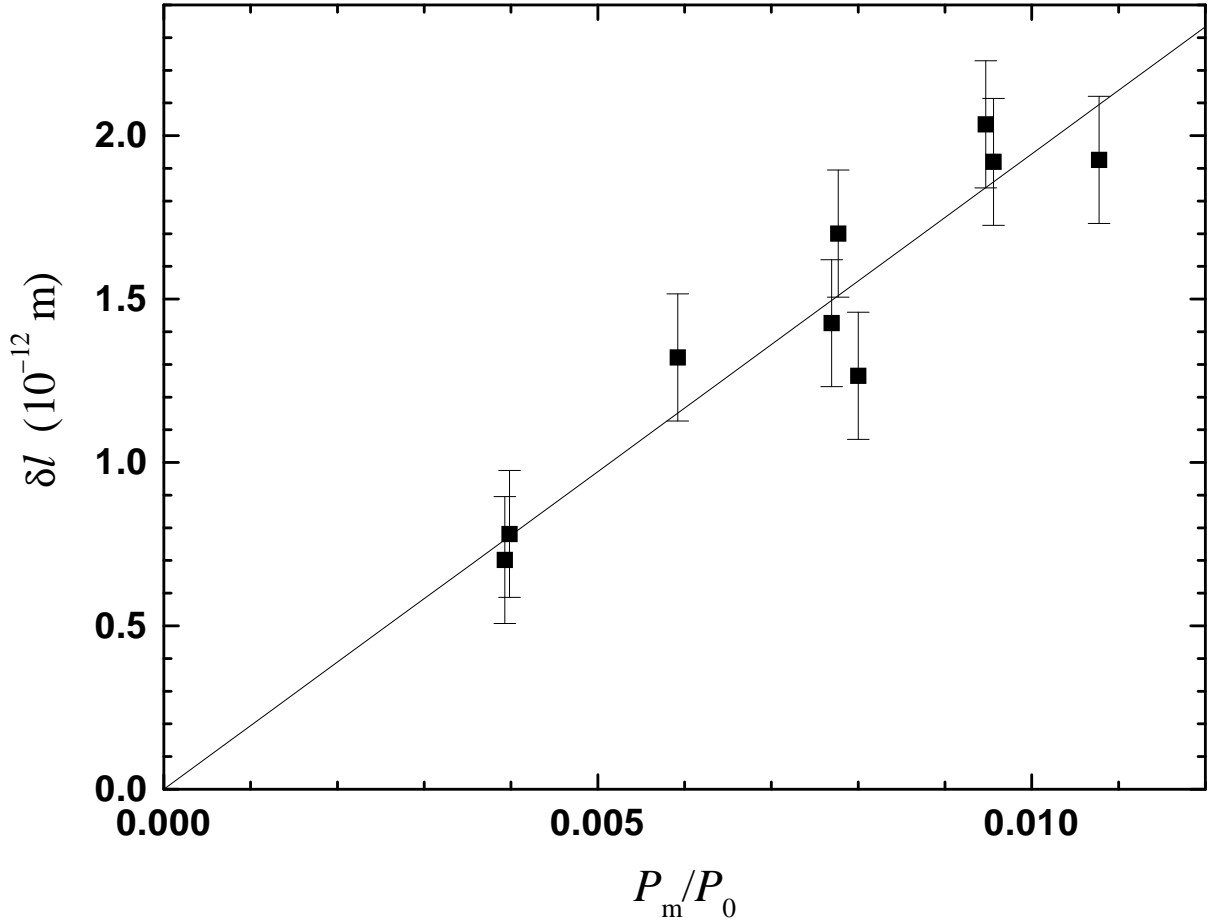} 
\end{center}
\caption{Differential length variation $\delta l$ as a function of the relative modulation amplitude $P_m/P_0$, for a fixed modulation frequency $\nu_m$ = 0.05~Hz. The error bars are due to the uncertainty of the slope of the Pound--Drever signal, which calibrates the length scale.}
\label{plot2}
\end{figure}

We have first checked that the observed effect is proportional to the modulation amplitude $P_m$ of power incident on the cavities. In Fig.~\ref{plot2} the experimental differential length changes $\delta l$ are plotted as a function of the relative modulation amplitude $P_m/P_0$, with $\nu_m=0.05$~Hz. The linear fit has a reduced $\chi^2$ of 0.75.

\begin{figure}[t]
\begin{center}
\includegraphics*[bbllx=50bp,bblly=45bp,bburx=690bp,bbury=530bp,width=\columnwidth]{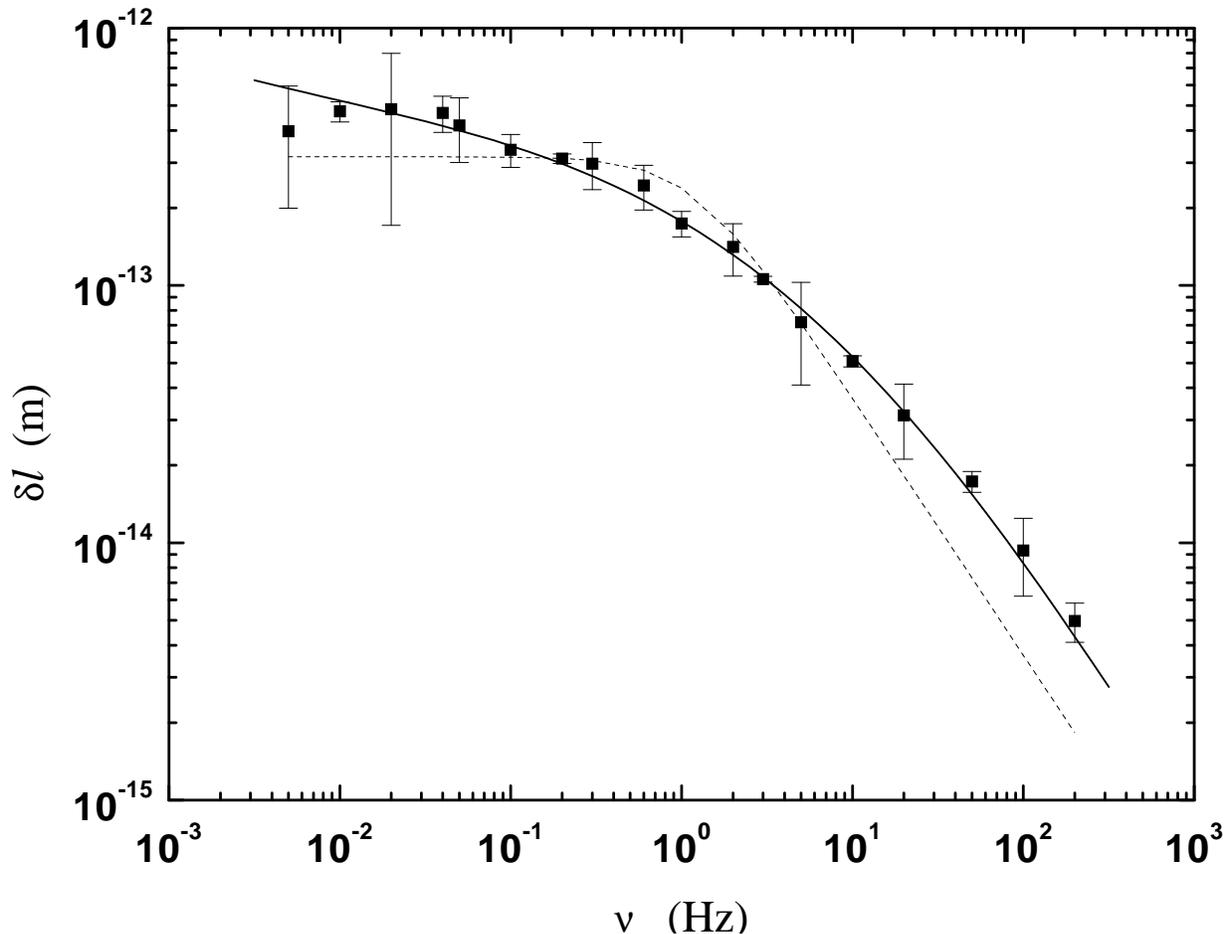} 
\end{center}
\caption{Differential length variation $\delta l$ as a function of the modulation frequency $\nu_m$. The solid line is the fitting curve of the experimental data with the expression of Eq.~(\ref{pinard}). The dashed line is the best fit of the experimental data with a single-pole, low-pass function.}
\label{plot1}
\end{figure}

In Fig.~\ref{plot1} we plot the experimental data for $\delta l$ as function of the modulation frequency $\nu_{\rm m}$. 
The reported error bars results from averaging over repeated acquisitions.
These data have been fitted according to the model presented in Ref.~\cite{Cerdonio}. From Eqs.~(\ref{eq:Sxx}) and (\ref{eq:integral}) we obtain: 
\begin{equation}
\delta l (\nu) = L_0 \sqrt{K(\nu/\nu_c)} \, ,
\label{pinard}
\end{equation}
where
\begin{equation} 
L_0 =  \frac{1}{\pi} \, (1+\sigma) \, \frac{ \alpha}{\kappa} \, \delta P_{\rm abs} \, ,
\label{eq:constant}
\end{equation}
and $\delta P_{\rm abs}$ is the difference in the modulated power absorbed by the mirrors of the two cavities. 
The resulting fitting curve, also displayed in Fig.~\ref{plot1}, shows an excellent agreement between theory and experimental data (reduced $\chi^2= 0.5$). 
>From the fit we get the scale factor $L_0=(3.3 \pm 0.4) \times 10^{-13}$~m and the cutoff frequency $\nu_c = 2.8 \pm 0.6$~Hz, where the reported error is the standard deviation. 
>From Eq.~(\ref{omega}) we calculate for $\nu_c=\omega_c/2\pi$ the values of 2.7~Hz, 2.3~Hz and 1.8~Hz using respectively the three beam waists reported in Table~\ref{tab1} and the following figures for fused silica at room temperature~\cite{Braginsky1}: $\kappa=1.4$~W/m~K, $\rho = 2.2 \times 10^{3}$~kg/m$^3$, $C=6.7 \times 10^2$~J/kg~K  
\footnote{Here we assume that the effect is mainly due to the substrate of the mirrors, even though multilayer coatings are presumed to play a role as well: unfortunately we do not know the exact composition of the coatings and their thermal and mechanical parameters.}.
We notice that the beam radius $r_0$, as defined above, is a factor $1/\sqrt{2}$ with respect to the mirror beam waist as usually defined in Gaussian optics, which we report in Table~\ref{tab1} for each mirror.
The agreement with the experiment is very good. To emphasize the peculiarity of the model we adopted, in Fig.~\ref{plot1} we also plot the best fitting curve (reduced $\chi^2=8$) of the experimental data with a single-pole, low-pass function, 
$\delta l = L_0/\sqrt{1+(\nu/\nu_c)^2}$. 
The functional form of $\delta l$ is a striking argument confirming the calculations of Cerdonio {\em et al.}

As far as the size of the effect is concerned, we actually ignore the fractional amount of power which is absorbed at each reflection. From the reflection and transmission coefficients of each cavity we calculate the mirror losses, however these losses are not only due to absorption but also to scattering. Anyhow, we can give an estimate of the mirrors absorption coefficient, assuming that it is roughly the same for all the four mirrors, and that the losses are mostly due to scattering. From the parameter $L_0$ resulting from the fit and Eq.~(\ref{eq:constant}), with $\alpha=5.5 \times 10^{-7}$~K$^{-1}$ and $\sigma=0.17$ for fused silica, we get $\delta P_{\rm abs} = 2.3 \times 10^{-6}$~W for the modulated absorbed power. The differential intracavity power $\delta P_c$ is 220~W, as from the value reported in Table \ref{tab1}, and about 1\% of it is actually modulated. Dividing the absorbed modulated power by the intracavity modulated power and by the number of mirrors (2) we obtain a mean value for the absorption coefficient of each mirror of about $5 \times 10^{-7}$, which is quite reasonable and in agreement with the estimate reported in Ref.~\cite{Braginsky1}. 

In conclusion we have experimentally investigated the photothermal effect on the effective length of high finesse  Fabry--Perot interferometers. The measurements have been performed by modulating the power impinging on two cavities. The differential length change was measured for different modulation depths and frequencies, by means of an actively stabilized  laser system with low residual frequency fluctuations and thanks to low noise, acoustically and thermally isolated cavities. 
The frequency functional dependence of the effect is perfectly explained by the model proposed by Cerdonio {\em et al.} and the experimental parameters obtained from the fitting procedure are in very good agreement with independent estimations.
Besides the interest for the physical phenomenon itself, this experimental study is a step toward a full understanding of thermal effects which are of critical importance for high sensitivity displacement measurements.


\begin{thebibliography}{99}

\bibitem{Saulson} P.~R.~Saulson,
Phys. Rev. D {\bf 42} (1990) 2437.

\bibitem{Levin} Yu.~Levin,
Phys. Rev. D {\bf 57} (1998) 659.

\bibitem{Braginsky1} V.~B.~Braginsky, M.~L.~Gorodetsky, and S.P.~Vyatchanin, 
Phys. Lett. A {\bf 264} (1999)~1.

\bibitem{Liu} Y.~T.~Liu and K.~S.~Thorne,
Phys. Rev. D {\bf 62} (2000) 122002.

\bibitem{Cerdonio} M.~Cerdonio, L.~Conti, A.~Heidmann and M.~Pinard,
Phys. Rev. D {\bf 63} (2001) 082003.

\bibitem{interferometri} A.~Giazotto, Phys. Rep. {\bf 182} (1989) 365.

\bibitem{Derosa} M.~De~Rosa {\em et al.}, 
Class. Quant. Grav. {\bf 19} (2002) 1919.

\bibitem{ligo2} LIGO document, M990288-A-M (1999); downloadable as a printable file at the LIGO web site: http://www.ligo.caltech.edu/docs/M/M990288-A1.pdf.

\bibitem{FEL}
S.~Benson, Michelle Shinn, and G.~R.~Neil,
in Proceedings of the 
22nd International Free Electron Laser Conference
and 7th FEL Users Workshop,Durham, North Carolina USA 2000,
edited by V.~N.~Litvinenko and Y.~K.~Wu.
The proceedings have been published on the Web and the cited contribution can be found at the following address:
http://www.fel.duke.edu/fel2000/program/proceedings/TH-2-04.pdf

\bibitem{MIT}
K.~An, A.~Sones, C.~Fang-Yen, R.~R.~Dasari, and M.~S.~Feld,
Opt. Lett. {\bf 22} (1997) 1433.

\bibitem{Conti-RSI} 
L.~Conti {\em et al.},
Rev. Sci. Instrum. {\bf 69} (1998) 554.

\bibitem{Conti-AO} L.~Conti, M.~De~Rosa and F.~Marin, 
Appl. Opt. {\bf 39} (2000) 5732.

\bibitem{Drever}
R.~W.~P.~Drever {\em et al.},
Appl. Phys. B {\bf 31} (1983) 97.
\end{thebibliography}
\end{document}